\documentclass[pre,aps,preprint,showpacs]{revtex4-1}
\usepackage{amsmath,amssymb,amstext,amscd,amsthm,amsopn,graphicx,indentfirst,mathrsfs}

\begin{document}

\title{Effects of non-resonant interaction in ensembles of phase oscillators}

\author{Maxim Komarov}
\affiliation{Faculty of Computational Mathematics and Cybernetics,
Nizhni Novgorod University, 23 Pr. Gagarina, 603950 Nizhni Novgorod, Russia}
\author{Arkady Pikovsky}
\affiliation{Department of Physics and Astronomy, Potsdam University,
  Karl-Liebknecht-Str 24, D-14476, Potsdam, Germany}

\date{\today}

\begin{abstract}
We consider general properties of groups of interacting oscillators, for which the natural
frequencies are not in resonance. Such groups interact via non-oscillating collective variables 
like the amplitudes of the order parameters defined for each group. We treat the phase
dynamics of the groups using the Ott-Antonsen ansatz and reduce it to a system of coupled 
equations for the order parameters. We describe different regimes of co-synchrony in the groups. 
For a large number of groups, heteroclinic cycles, corresponding to a sequental synchronous activity
of groups, and chaotic states, where the order parameters 
oscillate irregularly, are possible. 
\end{abstract}

\maketitle

\section{Introduction}

Models of coupled limit cycle oscillators are widely used to describe
self-synchronization phenomena in various branches of science. The applications 
include physical systems like 
Josephson junctions \cite{Wiesenfeld-Swift-95}, lasers \cite{Glova-03}, and 
electrochemical oscillators \cite{Kiss-Zhai-Hudson-02a}, but similar models are also used
for neuronal ensembles~\cite{Golomb-Hansel-Mato-01}, the dynamics of pedestrians
on bridges \cite{Strogatz_et_al-05,Eckhardt_et_al-07}, applauding 
persons~\cite{Neda-Ravasz-Brechet-Vicsek-Barabasi-00}, 
 etc.

In many cases a model of a fully connected (globally coupled) network is appropriate, 
it means that the oscillator population is treated in the 
mean field approximation. 
Ensembles of weakly interacting self-sustained oscillators are successfully handled in
the framework
 of phase approximation 
\cite{Kuramoto-75,Kuramoto-84,Daido-92a,Daido-93a,Daido-96}.
Most popular are the Kuramoto model of sine-coupled phase oscillators, and its extension, 
the Kuramoto-Saka\-gu\-chi model \cite{Sakaguchi-Kuramoto-86}.
This model describes self-synchroni\-za\-tion and appearance of a collective mode 
(mean field) in
an ensemble of generally non-identical elements as a nonequilibrium phase transition. 
The basic assumptions behind the Kuramoto model are that of weak coupling and of 
closeness of frequencies of oscillators, the latter results in the presence of resonant terms in the
coupling function only. References to detailed aspects of the
Kuramoto model can be found
in~\cite{Pikovsky-Rosenblum-Kurths-01,Acebron-etal-05,Strogatz-00}.

In many cases the ensembles of oscillators are not uniform and can be considered as
consisting of several subensembles (e.g., 
in brain different groups of neurons can have different characteristic rhythms). If one still assumes
that the frequencies of these subgroups are close (compared to the coupling), then a model
of several interacting subpopulations~\cite{Tukhlina-Rosenblum-08,Popovych-Tass-10} or of 
an ensemble having a bimodal (or a multi-modal)
distribution of frequencies~\cite{Bonilla-Neu-Spigler-92,Crawford-94,Bonilla_etal-98,Bonilla-00,%
Montbrio-Pazo-Schmidt-06,Martens_etal-09,Pazo-Montbrio-09} is adopted. 
Similarly, one can also model two ensembles, 
one of which consists of active and another of  passive elements, which are coupled resonantly 
due to closeness of their frequencies~\cite{Daido-Nakanishi-04,Pazo-Montbrio-06}.

In this paper we study a novel situation of \textit{non-resonantly} coupled oscillator ensembles.
We assume that there are several groups of oscillators, the frequencies in each group are close to each other, but are
strongly (compared to the coupling strength) different
between the groups. In this situation the coupling within the group is resonant, 
like in usual Kuramoto-type models, but the coupling between the groups can be only 
non-resonant~\footnote{Another novel type of interaction appears if the frequencies of two groups 
are in a high-order resonance like $2:1$, see~\cite{Luck-Pikovsky-11}.}. 
It means that the coupling can be via non-oscillating, slow variables only, i.e. via the 
amplitudes 
of the mean fields. In the context of a single Kuramoto model such a dependence 
on the amplitude of the mean field
corresponds to a 
nonlinearity of coupling, recently studied in~\cite{Rosenblum-Pikovsky-07,Pikovsky-Rosenblum-09,%
Filatrella-Pedersen-Wiesenfeld-07,Giannuzzi_et_al-07}.
Nonlinearity in this context means that the effect of the collective mode on an individual unit
depends on the amplitude of this mode, so that, e.g., the interaction of the field and of a 
unit can be attractive for a weak field and repulsive for a strong one. 
Mathematically, this is represented 
by the dependence of the parameters of the Kuramoto-Sakaguchi model 
(the coupling strength, the effective frequency spreading, and the phase shift) 
on the mean field amplitude. Here we generalize this approach to several ensembles, 
so that
the parameters of the  Kuramoto-Sakaguchi model describing each 
subgroup depend on the mean field amplitudes 
of other subgroups (e.g., resonant interactions within a group 
of oscillators can be attractive or repulsive dependent on the
amplitude of the order parameter of another group).

In Section~\ref{sec:bm} we introduce the basic model of non-resonantly interacting ensembles. We also formulate the equations for the 
mean fields of the ensembles following the Ott-Antonsen theory~\cite{Ott-Antonsen-08,Ott-Antonsen-09}. The simplest situation
of two interacting ensembles is studied in Section~\ref{sec:two}. In Section~\ref{sec:three} we describe  three and 
several interacting ensembles, focusing on
nontrivial regimes of sequential
synchronous activity following a heteroclinic cycle, and on chaotic dynamics.

\section{Basic model of non-resonantly interacting oscillator ensembles}
\label{sec:bm}
\subsection{Kuramoto-Sakaguchi model and Ott-Antonsen equations for its dynamics}
A popular model describing \textit{resonant} interactions in an ensemble of oscillators having close frequencies
is due to Kuramoto and Sakaguchi~\cite{Sakaguchi-Kuramoto-86}
\begin{equation}
 \dot \phi_k=\omega_k+\text{Im}(K Z e^{-i\phi_k}),\qquad Z=\frac{1}{N}\sum e^{i\phi_k},\qquad k=1,\ldots,N.
\label{eq:kursak}
\end{equation}
Here $\phi_k$ is oscillator's phase, $Z$ is the complex order parameter (mean field)
that also serves as a measure for synchrony in the ensemble, $\omega_k$ 
are natural frequencies of oscillators, and $K=2a+2ib$ is a complex coupling constant.
Recently, Ott and Antonsen~\cite{Ott-Antonsen-08,Ott-Antonsen-09} have demonstrated that 
in the thermodynamic limit $N\to\infty$, and asymptotically for large times
the evolution of the order parameter $Z$ in the case of a Lorentzian distribution 
of natural frequencies  $g(\omega)=\Delta [\pi(\omega-\omega_0)^2+\Delta^2]^{-1}$
around the central frequency $\omega_0$
is governed by a simple ordinary differential equation
\begin{equation}
 \dot Z=(i\omega_0-\Delta)Z+\frac{1}{2}(K-K^*|Z|^2)Z.
\label{eq:oa}
\end{equation}
Written for the amplitude and the phase of the order parameter defined according to $Z=\rho e^{i\Phi}$,
the Ott-Antonsen equations 
\begin{align}
 \dot\rho&=-\Delta \rho+a(1-\rho^2)\rho\;,
\label{eq:oareal}\\
\dot\Phi&=\omega_0+2b\rho^2\;,
\label{eq:oaimag}
\end{align}
are easy to study: Eq.~(\ref{eq:oareal}) defines the stationary amplitude of the mean field 
(which is non-zero above the synchronization threshold $a_c=\Delta$), while Eq.~(\ref{eq:oaimag})
yields the frequency of the mean field. 

\subsection{Non-resonantly interacting ensembles}
We consider several ensembles of oscillators, each characterized by its own parameters
$\omega_0,\Delta,a,b$. The main assumption is that the central frequencies $\omega_0$ of different 
populations are not close to each other, and also high-order resonances between them are not present.
Such a situation appears typical for neural ensembles, where different areas of brain demonstrate
oscillations in a very broad range of frequencies, from alpha to gamma rhythms.
Because there is no resonant interaction between the oscillators in different ensembles, they can interact
only non-resonantly, via the absolute values of the mean fields. Assuming 
that only Kuramoto order parameters (\ref{eq:kursak}) (but not higher-order Daido order 
parameters $Z_m=\langle e^{im\phi}\rangle$) 
enter the coupling, a general non-resonant
interaction between populations can be described by the dependencies of the parameters 
$\omega_0,\Delta,a,b$ on the amplitudes of the mean fields $\rho_l$, where index $l$ counts the
subpopulations. 
Moreover, one can see from  (\ref{eq:oareal},\ref{eq:oaimag}) that the equation
for the amplitude is independent on the phase, therefore we can restrict our attention to the amplitude
dynamics (\ref{eq:oareal}). Furthermore, we assume the coupling to be week, so only the leading order
corrections $\sim\rho^2$ are 
included. All this leads to the following general model for interacting
populations
\begin{equation}
 \dot \rho_l=(-\Delta_l-\Gamma_{lm}\rho_m^2)\rho_l+(a_l+A_{lm}\rho_m^2)(1-\rho_l^2)\rho_l,\qquad l=1,\dots,L
\label{eq:basmod}
\end{equation}
with coupling constants $\Gamma_{lm},A_{lm}$. Note that because the widths 
of the frequencies distribution 
cannot be negative, coefficients $\Gamma_{lm}$ must satisfy $\Delta_l+\Gamma_{lm}\geq 0$. Below we assume
that there is no nonlinearity inside ensembles $\Gamma_{ll}=A_{ll}=0$.

In this paper we will not investigate model (\ref{eq:basmod}) in its full generality, as it would require a rather 
tedious analysis. Instead, we will consider two simpler models, which describe particular types 
of interaction, but nevertheless allow us to demonstrate interesting dynamical patterns.
In model A we assume that only frequencies are influenced by the
coupling, i.e. $A_{lm}=0$. This leads
to a system
\begin{equation}
 \dot \rho_l=(a_l-\Delta_l-\Gamma_{lm}\rho_m^2-a_l\rho_l^2)\rho_l
\label{eq:modelA}
\end{equation}
Another model B takes into account the interaction via coupling constants only (i.e. $\Gamma_{lm}=0$); 
additionally
we will assume here that the distributions of frequencies in all interacting ensembles are narrow $\Delta_l\to0$.
In the limit of identical oscillators we obtain from (\ref{eq:basmod})
\begin{equation}
  \dot \rho_l=(a_l+A_{lm}\rho_m^2)(1-\rho_l^2)\rho_l
\label{eq:modelB}
\end{equation}
Here we note that the Ott-Antonsen equations for the ensemble of identical oscillators describe not a general case, but a particular
solution, while a general description delivers the 
Watanabe-Strogatz theory~\cite{Watanabe-Strogatz-94,Pikovsky-Rosenblum-08}. Thus the dynamics of 
model B should be considered as a special singular limit $\Delta\to0$.
 
Below, in sections \ref{sec:two} and \ref{sec:three} we describe the dynamics of these two models, for the cases
of two, and three and more interacting 
ensembles, respectively.

\section{Two interacting ensembles}
\label{sec:two}
Let us first rewrite models (\ref{eq:modelA},\ref{eq:modelB}) for the simplest case of only two 
interacting ensembles. Additionally, for model A we assume $a_l=1$ (equivalently, one could
 renormalize the amplitudes of order parameters 
$\rho_{1,2}$ to get rid of these coefficients). Thus the model A reads
\begin{equation}
 \begin{aligned}
 \dot\rho_1&=\rho_1(\delta_1 -d_{12}\rho_2^2-\rho_1^2)\;,\\
\dot\rho_2&=\rho_2(\delta_2 -d_{21}\rho_1^2-\rho_2^2)\;.
\end{aligned}
\label{eq:modelA2}
\end{equation}
For model B a 
normalization of amplitudes is not possible, and it reads
\begin{equation}
\begin{aligned}
 \dot\rho_1&=\varepsilon_1\rho_1(1-D_{12}\rho_2^2)(1-\rho_1^2)\;,\\
\dot\rho_2&=\varepsilon_2\rho_1(1-D_{21}\rho_1^2)(1-\rho_2^2)\;.
\end{aligned}
\label{eq:modelB2}
\end{equation}
Generally, parameters $\delta_{1,2}=1-\Delta_{1,2},\;d_{ik}=\Gamma_{ik},\;\varepsilon_{1,2}=a_{1,2},\;D_{ik}=-A_{ik}/a_i$ can have different signs.

As the first property of both models we mention that the dynamics is restricted to the domain $0\leq \rho_{1,2}\leq 1$.
Formally, this follows directly from (\ref{eq:basmod}), physically this corresponds to the admissible range of values of 
the order parameter. Furthermore, for model A (\ref{eq:modelA2}) we can apply the Bendixon-Dulac criterion
$$
\frac{\partial}{\partial\rho_1}\left(\frac{1}{\rho_1\rho_2}\dot{\rho}
_1\right)+\frac{\partial}{\partial\rho_2}\left(\frac{1}{\rho_1\rho_2}\dot{\rho}
_2\right)=-2\frac{\rho_1^2+\rho_2^2}{\rho_1\rho_2}<0  
$$
from which it follows that it cannot possess periodic orbits.  

Remarkably, model B (\ref{eq:modelB2}) can be written as a Hamiltonian one.
With an ansatz 
\begin{equation}
 \exp y_{1,2}=\rho_{1,2}^2(1-\rho_{1,2}^2)^{-1}
\label{eq:hamans}
\end{equation}
it can be represented in a Hamiltonian form
\begin{equation}
\begin{gathered}
\dot y_1=\frac{\partial H(y_1,y_2)}{\partial y_2},\qquad \dot y_2=-\frac{\partial H(y_1,y_2)}{\partial y_1},\\
\qquad H=2\varepsilon_1 y_2-2\varepsilon_2y_1-2\varepsilon_1D_{12}\ln(1+e^{y_2})+2\varepsilon_2 D_{21}\ln(1+e^{y_1})\;.
\end{gathered}
\label{eq:hamilt}
\end{equation}
Thus model B  may demonstrate a family of periodic orbits if the levels
of the Hamiltonian are closed curves. We stress that the Hamiltonian structure of the model does not exclude
existence of stable equilibria at $\rho=0,1$ because the transformation (\ref{eq:hamans}) is singular at these states;
in the Hamiltonian formulation (\ref{eq:hamilt}) these stable equilibria correspond to trajectories moving toward $\mp \infty$.

The dynamics of both models is mainly determined by the existence and stability 
of equilibria. For model A (\ref{eq:modelA2}) possible equilibria are the trivial one $S_1(0,0)$, two states where
one of the order parameters vanish $S_2(\delta_1^{1/2},0)$ and $S_3(0,\delta_2^{1/2})$, 
and a state where both order parameters are non-zero
$S_4((\delta_1-d_{12}\delta_1)(1-d_{12}d_{21})^{-1},(\delta_2-d_{21}\delta_1)(1-d_{12}d_{21})^{-1})$.  
Similarly, model B (\ref{eq:modelB2}) 
always has equilibria $M_1(0,0)$, $M_2(1,0)$, $M_3(0,1)$ and $M_4(1,1)$, and 
additionally a nontrivial state $M_5(D_{21}^{-1/2},D_{12}^{-1/2})$ existing if $D_{12},D_{21}>1$.

\begin{figure}
\centering
\includegraphics[width=\columnwidth]{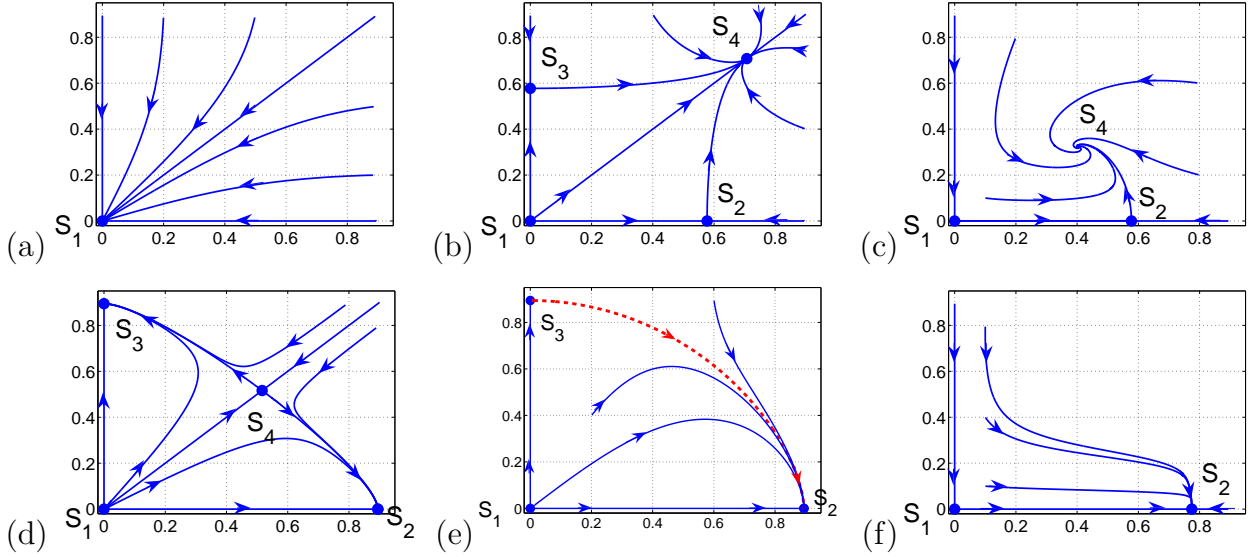}
\caption{(Color online) Six different patterns of the dynamics of system (\ref{eq:modelA2}). 
(a) Global stability of a trivial state (for $\delta_{1,2}<0$). 
(b,c) Global stability of $S_4$ when both populations are
partially synchronous (conditions for this are (\ref{case2a-1}) for (b) or
(\ref{case2a-2}) for (c)). (d) Competition between clusters if the coupling
is strongly suppressive (\ref{case3a})); 
here we have bistablity of states $S_{2,3}$ describing 
 synchronous activity of one cluster and asynchronous of
another one. (e) Asymmetric interaction between clusters arises under condition 
(\ref{case4a-2}); here always a heteroclinic
trajectory from saddle point $S_3$ to stable node $S_2$ exists (red dashed line). (f) Global stability of
$S_2$ under condition (\ref{case4a-2}).}
\label{modes}
\end{figure}

\begin{figure}
\centering
\includegraphics[width=\columnwidth]{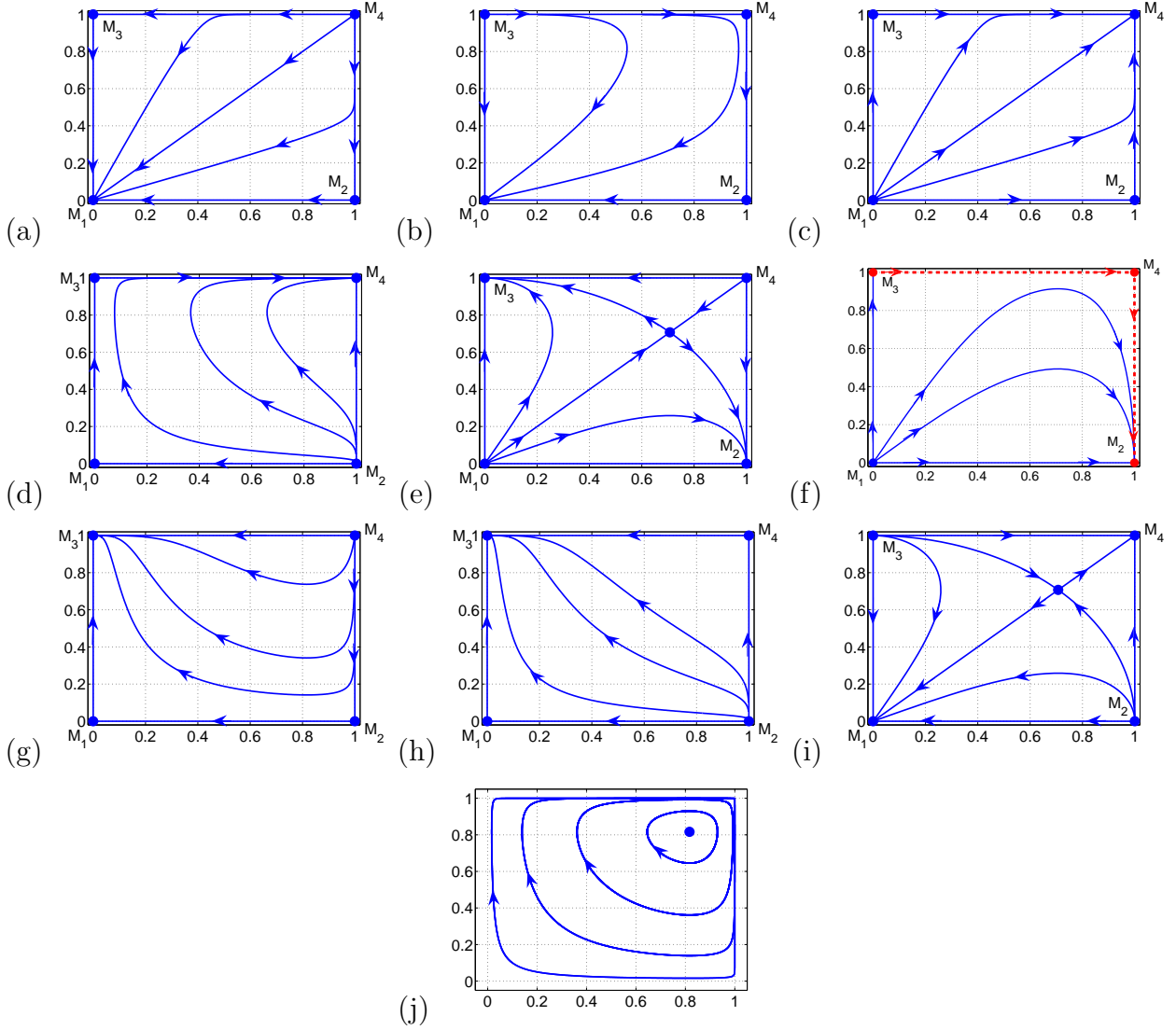}
\caption{(Color online) Ten different dynamical regimes in system (\ref{eq:modelB2}). 
(a,b): Global stability of the trivial state, arises at conditions (\ref{case1b}). 
Case (a): $D_{12}<1$, case (b): $D_{12}>1$. (c,d):
Global stability of $M_4(1,1)$ when both clusters are in the 
synchronized state,
under condition of a weak suppressive coupling (\ref{case2b-1})
for (c) or at (\ref{case2b-2}) for (d). 
(e): Competition between clusters, 
arises at strong suppressive coupling (\ref{case3b}). Here
we have bistablity of steady states $M_{2,3}$; each of these points 
corresponds to
synchronous activity of one cluster and asynchronous of another one. 
(f): Asymmetric interaction between clusters at  asymmetric coupling (\ref{case4b-2}). 
Here always a sequence of
heteroclinic trajectories $M_3\to M_4\to M_2$  (red dashed
lines) is present. (g,h): The situation of global stability of fixed point $M_3$
while conditions (\ref{case4b-1}) are satisfied (case (g): $D_{21}>1$, case (h): $D_{21}<1$). 
(i): Bistablity of fully asynchronous and fully synchronous states, arises if (\ref{case5b}) is valid. 
In this case stable manifolds of the saddle point $M_5$ divide basins of attraction of
stable points $M_1$, $M_4$. (j): The case of periodic behavior,  arises at
conditions (\ref{case6b}).}
\label{modes1}
\end{figure}

We illustrate possible types of dynamics (up to symmetry $1\leftrightarrow 2$) in models A,B in 
Figs.~\ref{modes},\ref{modes1}. Here it is worth mentioning, that model (\ref{eq:modelA2}) is
structurally of the same type as typical models of interacting populations in 
mathematical ecology~\cite{Murray-02}.
Model B  (\ref{eq:modelB2}) resembles them as well, but has a distinctive property that fully synchronized cluster 
$\rho=1$ is invariant. Referring for the details to Appendix~\ref{appendix1}, we describe 
briefly possible regimes in these models.

\begin{enumerate}

\item Global stability of trivial equilibrium point $S_1(0,0),\
M_1(0,0)$ (Fig.\ref{modes}a, Fig.\ref{modes1}a,b) means that a fully asynchronous state
is stable in both ensembles. 
\item Stability of a nontrivial state off coordinate axes $S_4$ and
$M_4$ (Fig.\ref{modes}b,c, Fig.\ref{modes1}c,d). Here both ensembles are synchronized
(in model A not completely because of a distribution of frequencies, in model B completely because we 
assume identical oscillators in ensembles).

\item Competition between ensembles
(Fig.\ref{modes}d,\ref{modes1}e): Only one ensemble synchronizes while 
the other one desynchronizes. Which ensemble is synchronous depends on initial conditions.

\item Suppression:  One ensemble always ``wins'' and is synchronous while the other one desynchronizes
(steady states $S_2$,$M_2$ are global attractors, of course also stability of ``symmetric'' states $S_3,M_3$ 
is possible)(Fig.\ref{modes}e,f,\ref{modes1}f,g,h).

\item The case of bistability of the  trivial and the fully synchronous states of both ensembles
(Fig.\ref{modes1}i) is possible in the model B only. 

\item Periodic behavior (Fig.\ref{modes1}j) is possible only in ensemble B, it corresponds to an 
interaction of populations of ``predator-pray'' type. Because of the system is Hamiltonian,
the oscillations are conservative like in the Lottka-Volterra system.
\end{enumerate}

\begin{figure}
\centering
\includegraphics[width=\columnwidth]{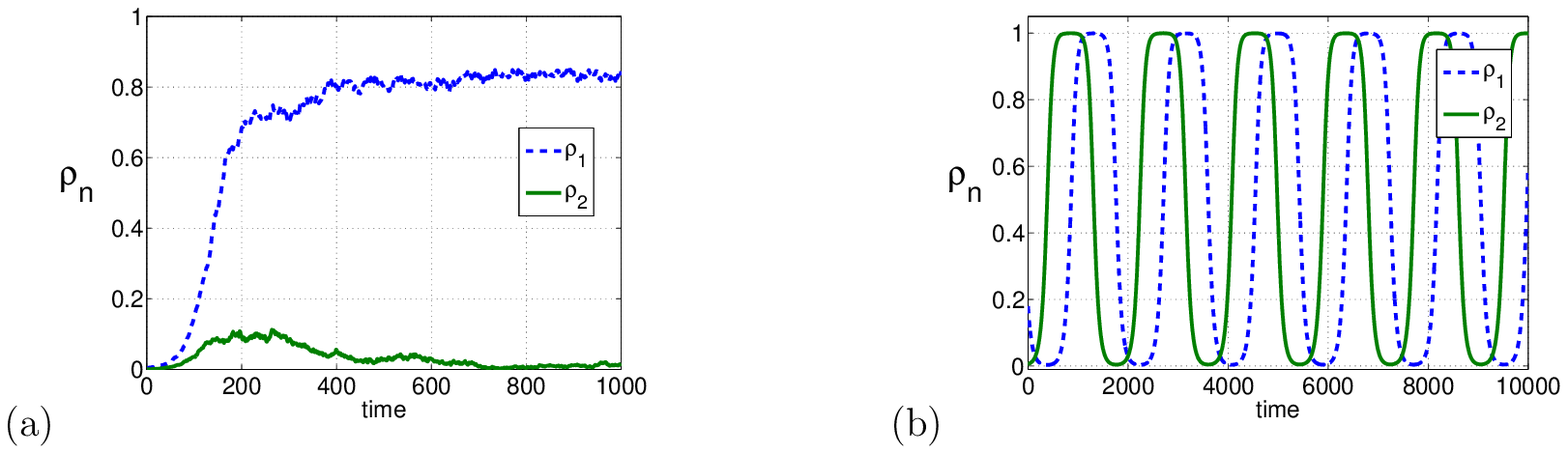}
\caption{(Color online) Modeling of ensemble consisting of two subpopulations of $N=10^3$ 
phase oscillators. (a) Subpopulations 
interact via modulation of effective frequency mismatch~(\ref{eq:modelA2}). 
Case of competition between subpopulations for
parameter values $\delta_{1,2}=10$,  $d_{12}=d_{21}=12$. 
(b) Subpopulations interact via coupling modulations~(\ref{eq:modelB2}). 
A periodic regime is presented at 
parameter values $\varepsilon_1=-1$, $\varepsilon_2=1$, $D_{12}=D_{21}=2$. 
To avoid a spurious clustering and to ensure validity of Ott-Antonsen description, a 
small mismatch was added: $\omega_n$ 
were randomly distributed in the range $[-0.025,+0.025]$.}
\label{fig:initial}
\end{figure}

While in our analysis we studied models (\ref{eq:modelA2},\ref{eq:modelB2}) describing dynamics of the order parameters in the
Ott-Antonsen ansatz, all the regimes
 described above can be observed when one simulates original equations of the ensembles of 
phase oscillators~(\ref{eq:kursak}),  at sufficiently large number of units $N$. 
In Fig.~\ref{fig:initial} we illustrate two nontrivial regimes of two subpopulations of phase 
oscillators at $N=10^3$. 
Figure \ref{fig:initial}(a) shows the dynamics of mean fields in the case of a 
competition between two subpopulations that interact via frequency mismatch modulation, see Fig.~\ref{modes}(d). 
Figure \ref{fig:initial}(b) illustrates a periodic behavior of two subpopulations like in Fig.~\ref{modes1}(j).

\section{Three and more interacting ensembles}
\label{sec:three}
In this section we generalize the results of Section~\ref{sec:two} to many interacting ensembles. 
We do not aim here at the full generality, but rather present interesting regimes based on the elementary
dynamics depicted in Figs.~\ref{modes},\ref{modes1}. According to the consideration above, we restrict
our attention to 
two basic models A (\ref{eq:modelA}) and B (\ref{eq:modelB}). Generally, model B cannot be rewritten in a 
Hamiltonian form, but by applying transformation (\ref{eq:hamans}) one can easily see that this system has a Liouvillian property -- 
the phase volume is conserved.

\subsection{Symmetric case: cosynchrony and competition}
\begin{figure}
\centering
\includegraphics[width=\columnwidth]{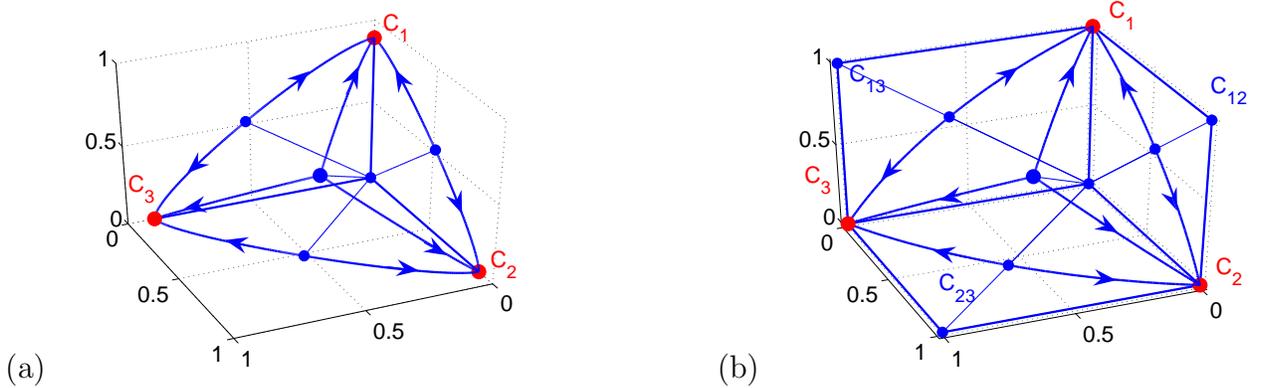}
\caption{(Color online) Multistability of steady states $C_n$ 
corresponding to synchronous state of only one cluster for (a) system (\ref{eq:modelA}) 
($\Delta>a$,\ $\Gamma>a$) and (b) system (\ref{eq:modelB}) 
($a>0$, $A<-a$).}
\label{fig:3mult}
\end{figure}

Here we describe mostly simple regimes that are observed in a symmetric case where parameters
of all ensembles and their interaction are equal. This corresponds to equal values $a_l=a, \Delta_l=\Delta,
\Gamma_{lm}=\Gamma$ in (\ref{eq:modelA}) and $a_l=a, A_{lm}=A$ in  (\ref{eq:modelB}). In model A, the only nontrivial regimes are
those where asynchronous states are unstable $\Delta<a$. Then one observes either a coexistence of synchrony like 
in Fig.~\ref{modes}b (for $\Gamma<a$) or a competition like in Fig.~\ref{modes}d (for $\Gamma>a$). In the latter case
only one 
ensemble is synchronous, while other desynchronize. Similar regimes can be observed in model B for $a>0$, $A<-a$.
 Additionally, in model B a coexistence of full synchrony in all ensembles and a full asynchrony, like in 
Fig.~\ref{modes1}i can be observed for $a<0$, $A>-\frac{a}{L-1}$. We illustrate the regimes of competition in Fig.~\ref{fig:3mult} for the case of
three interacting populations.

\subsection{Heteroclinic synchrony cycle}

Here we discuss a multidimensional generalization of the interactions where in a pair of ensembles
one group always synchronizes while another one is asynchronous (see Figs.~\ref{modes}(e),\ref{modes1}(f)). In the examples
presented in these graphs, both ensembles 
would self-synchronize separately, but due to interaction synchrony in ensemble 2 disappears while
ensemble 1 remains synchronous. One can say that in synchrony competition
between the first and the second ensembles,  the first ensemble wins. 
Suppose now, that a third self-synchronizing ensemble is added, which wins in the 
competition with the first one 
but looses in the competition to the second one. 
Then a cycle $2\to 1\to 3\to 2\to 1\to 3\ldots$ will be observed. Moreover, because
in the dynamics   Figs.~\ref{modes}(e),\ref{modes1}(f) the transition $2\to 1$ follows the heteroclinic orbit connecting 
steady states $S_3$ and $S_2$, the cycle in the system of three ensembles will be a heteroclinic one, with asymptotically
infinite period. Such a cycle has been studied in different 
contexts~\cite{Busse-Clever-79,Clune-Knobloch-94,Guckenheimer-Holmes-88}. For a review of 
robust heteroclinic cycles see \cite{Krupa-97,DynSys-special-issue-10} (sometimes one uses a term
``winnerless competition'' to describe such a dynamics~\cite{Rabinovich_etal-01,Afraimovitch_etal-08}).

\begin{figure}
\centering
\includegraphics[width=\columnwidth]{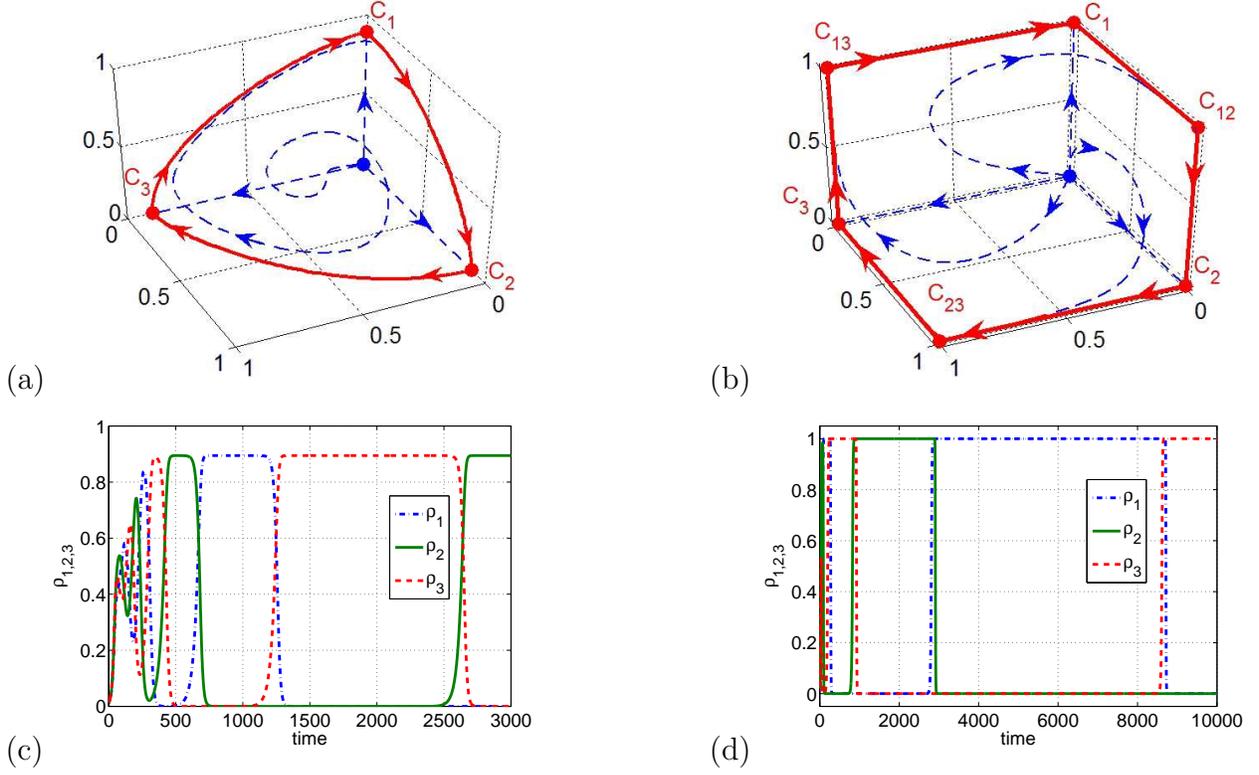}
\caption{(Color online) Stable heteroclinic cycles caused by asymmetric interactions between clusters 
in system (\ref{eq:modelA}) (a,c) and in system (\ref{eq:modelB}) (b,d). 
Parameters: (a,c) $a_l-\delta_l>0$, $\Gamma_{12}>\frac{a_2(a_1-\delta_1)}{a_2-\delta_2}$, $\Gamma_{31}>\frac{a_1(a_3-\delta_3)}{a_1-\delta_1}$, 
$\Gamma_{23}>\frac{a_3(a_2-\delta_2)}{a_3-\delta_3}$, $\Gamma_{21}<\frac{a_1(s_2-\delta_2)}{a_1-\delta_1}$, $\Gamma_{13}<\frac{a_3(a_1-\delta_1)}{a_3-\delta_3}$, $\Gamma_{32}<\frac{a_2(a_3-\delta_3)}{a_2-\delta_2}$, (d) $a_l>0$, $A_{12}<-a_1,\ A_{31}<-a_3,\ A_{23}<-a_2$, $A_{21}>-a_2,\ A_{13}>-a_1,\ A_{32}>-a_3$. 
Panels (a,b) show the phase space portraits while time series are presented in panels (c,d).}
\label{fig:3het}
\end{figure}

We demonstrate the heteroclinic synchrony cycle for three interacting ensembles in Fig.~\ref{fig:3het}. One can 
see that synchronous states of ensembles appear for longer and longer time intervals.
It is interesting to note that heteroclinic cycles have been observed in ensembles 
of \textit{identical} coupled oscillators
\cite{Ashwin92,Hansel93,Kori-Kuramoto-01,Kori-03,Ashwin-Borresen-04,Ashwin-Borresen-05}. 
There the nontrivial dynamics is
in the switchings of full synchrony between different clusters. In this respect the heteroclinic cycle in the model B
resembles such a regime. On the other hand, the heteroclinic cycle in model A is different: here 
the natural frequencies of oscillators are different and the states of synchrony are not complete, so the identical 
clusters never appear.

\begin{figure}
\centering 
\includegraphics[width=\columnwidth]{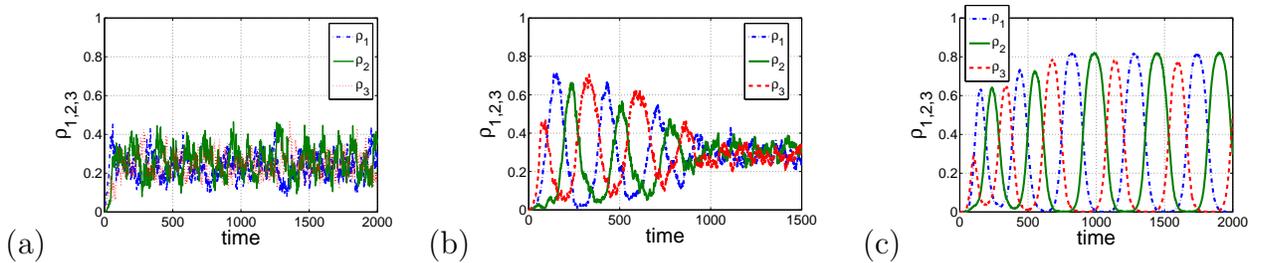}
\caption{(Color online) Dynamics of the order parameters of three interacting populations of oscillators 
(parameters like in Fig.~\ref{fig:3het} (a,c))
for three different sizes of populations: (a) $N=100$, (b) $N=400$, and (c) $N=10000$.}
\label{fig:ts-metastable}
\end{figure}

Finite size effects are nontrivial for the heteroclinic cycles described. Indeed, it is known that while in the thermodynamic
limit deterministic equations for the order parameters can be used, finite size effects can be modeled 
via noisy terms that scale roughly
as $\sim N^{-1/2}$~\cite{Pikovsky-Ruffo-99,Pikovsky-Zaikin-Casa-02,Hildebrand-Buice-Chow-07}.
On the other hand, noisy terms destroy perfect heteroclinic orbit, making the transitions times between the states finite and 
irregular. Exactly this is observed at modeling the interacting finite size ensembles (Fig.~\ref{fig:ts-metastable}). 
While for small $N$ the heteroclinic
cycle is completely destroyed, for large $N$ it looks like a noisy limit cycle.

\subsection{Chaotic oscillations}

\begin{figure}
\includegraphics[width=\columnwidth]{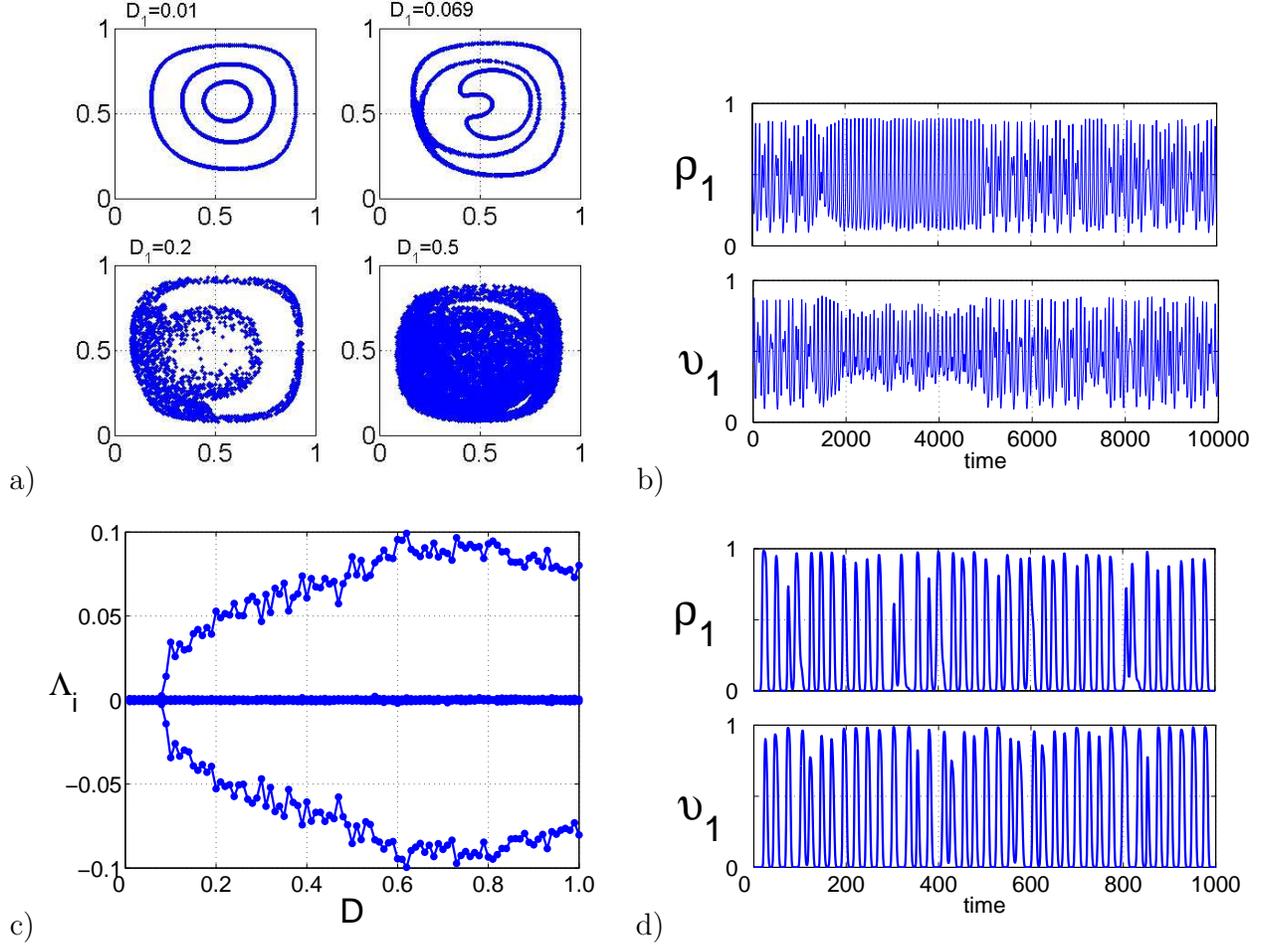}
\caption{ (a) 
Poincar\'e sections on the plane $(\upsilon_1,\upsilon_2)$ demonstarting regular and chaotic
dynamics at 
different values of $D_1$ in the system (\ref{eq:modelB-ch}). 
(b) Time series of a chaotic regime of system (\ref{eq:modelB-ch}), for 
parameter values  $D_1=0.5$, $D_0=2.0$, $\varepsilon_1 = -1.0$, $\varepsilon_2 = 1.0$. 
(c) Lyapunov exponents calculated at different values of $D_1$, for some particular value of the Hamiltonian. 
From four Lyapunov exponents two always vanish, while other two vanish for small $D_1$ (quasiperiodicity)
and are non-zero for larger couling (chaos).
(d) Chaotic time series of order parameters  of four subpopulations of oscillators 
consisting of $N=10^3$ elements each (the coupling
configuration and the parameters are like in panel (b)).}
\label{fig:ts-chaos}
\end{figure}

Here we discuss possible ``predator-pray''-type regimes (cf. Fig.~\ref{modes1}j) for many ensembles.
An elementary ``oscillator'' depicted in  Fig.~\ref{modes1}j can be represented 
as a Hamiltonian system with one degree of freedom.
Several of such elementary 
conservative ``oscillators'', being coupled, can yield quasiperiodic and chaotic regimes. In the case 
of two interacting conservative ``oscillators'' (i.e. of four interacting ensembles),  
system (\ref{eq:modelB}) can be rewritten as follows:
\begin{equation}
\begin{aligned}
\dot\rho_{1,2}&=\varepsilon_{1,2}\rho_{1,2}(1-D_{0}\rho_{2,1}^2 - D_{1}\upsilon_{1,2}^2)(1-\rho_{1,2}^2)\;,\\
\dot\upsilon_{1,2}&=\varepsilon_{1,2}\upsilon_{1,2}(1-D_{0}\upsilon_{2,1}^2 - D_{1}\rho_{1,2}^2)(1-\upsilon_{1,2}^2)\;.
\end{aligned}
\label{eq:modelB-ch}
\end{equation}
Here the parameters of the system were chosen in such a way that each pair of 
subpopulation $(\rho_{1},\rho_{2})$ and $(\upsilon_{1},\upsilon_2)$ exhibits 
periodic oscillation being decoupled from another pair (at $D_1=0$), i.e. $\varepsilon_1\varepsilon_2<0$ 
and $D_0>1$. When the coupling between the two pairs is introduced (i.e. $D_1\neq 0$),
then in dependence on this coupling and initial conditions the dynamics 
can be qusiperiodic or chaotic. Like in general Hamiltonian systems with two degrees of freedom, it
is convenient to represent the dynamics as a two-dimensional Poincar\'e  map.
As a Poincar\'e section (Figure \ref{fig:ts-chaos}a) we have taken the plane $(\upsilon_1,\upsilon_2)$ at
moments of time at which the variable  $\rho_1(t)$ has a maximum.  
At small values of the coupling 
between the ``oscillators'' $D_1$ the dynamics is typically quasiperiodic. While 
increasing $D_1$, one can observe a transition to dominance of chaotic regimes in the 
system (\ref{eq:modelB-ch}) (see Fig.\ref{fig:ts-chaos}a,b and calculation of 
Lyapunov exponents in Fig.\ref{fig:ts-chaos}c). 
Furthermore, we have confirmed the existence of chaotic 
oscillations by direct numerical simulation of four subpopulations satisfying~(\ref{eq:modelB-ch}), 
consisting of $N=10^3$ elements each (Fig.\ref{fig:ts-chaos}d). 

\section{Conclusion}
In this paper we have introduced and studied a model of non-resonantly coupled ensembles of oscillators. It is assumed that
oscillators form several groups, in each group the natural frequencies are close to each other, but the frequencies 
of different groups are rather different. This means that only oscillators within each group interact resonantly 
(i.e. the coupling terms depend on their phases), while interactions between the groups can be only non-resonant, i.e. depending on
slow non-oscillating variables only. As a particular realization of such a setup we considered phase oscillators, which resonantly interact
according to the Kuramoto-Sakaguchi model, and the non-resonant terms appear as dependencies of the 
parameters of the Kuramoto-Sakaguchi
model on the amplitudes of the mean fields (Kuramoto order parameters) of other groups. 

We employed the Ott-Antonsen 
theory allowing us to write a closed system of equation for the amplitudes of the order parameters. Analysis of this system constitutes 
the main part of the paper. The system resembles the Lottka-Volterra type equations used in mathematical ecology for the 
dynamics of populations, but has nevertheless some peculiarities. 
For two coupled ensembles we demonstrated a variety of possible regimes: coexistence
and bistability of synchronous states, as well as periodic oscillations. For a larger number of interacting
groups more complex states appear:
a stable heteroclinic cycle and a chaotic regime. Heteroclinic cycle means a sequence of synchronous epochs that become 
longer and longer. In a chaotic regime the order parameters demonstrate low-dimensional chaos. 
While the main analysis is performed for 
the Ott-Antonsen equations that are valid in the thermodynamic limit of infinite number of oscillators 
in ensembles, we have checked finite-size effects in several regimes 
by modeling finite ensembles. Finiteness of ensembles only slightly influences the dynamics 
in most of the observed states, except 
for the heteroclinic cycle. Here a small effective noise due to finite-size effects destroys the cycle, producing nearly periodic 
noise-induced oscillations.

One of the models we studied was that of groups of identical oscillators. 
Here in many cases only the states where some groups completely synchronize 
(i.e. all oscillators form an identical cluster) while other completely desynchronize (order parameter vanish)
are possible. Heteroclinic cycle in this model also connects such  states. There is, however, a nontrivial set of parameters, 
at which the order parameters of ensembles oscillate between zero and one, thus demonstrating time-dependent 
partial synchronization. Moreover, for four ensembles these oscillations are chaotic. This regime is quite interesting
for a general theory of collective chaos in oscillator populations 
(cf. chaotic dynamics of the order parameter in an ensemble of Josephson 
junctions reported in~\cite{Watanabe-Strogatz-94}) and certainly deserves further investigation. 

\begin{acknowledgments}
We thank M. Rosenblum, V. Petrov, G. Osipov and G. Bordyugov for 
useful discussions. M.K. acknowledges support from German-Russian Interdisciplinary Science Center (G-RISC) funded by the German Federal Foreign Office via the German Academic Exchange Service (DAAD), support from the Federal Programm (contracts No $\Pi15$, $\Pi2308$, $\Pi942$, 02.740.11.5138, 02.740.11.0839, 14.740.11.0348) and from the Russian Fund of Basic Research (08-02-92004, 08-02-970049, 10-02-00940) 
\end{acknowledgments}

\appendix
\section{Details of analysis of two interacting ensembles}
\label{appendix1}

Here we present details of the analysis of models (\ref{eq:modelA2},\ref{eq:modelB2}), giving the
conditions for different regimes presented in Figs.~\ref{modes}, \ref{modes1}.
\begin{enumerate}
\item The case of global stability of trivial equilibrium point $S_1(0,0),\
M_1(0,0)$ (Fig.\ref{modes}a, Fig.\ref{modes1}a,b).  For system (\ref{eq:modelA2}) 
such a situation occurs in the
case  $\delta_{1,2}<0.$  For system (\ref{eq:modelB2}) global stability of trivial
state $M_1$ occurs if $\varepsilon_{1,2}$ are negative and
at least one of $D_{12}$ or $D_{21}$ less than 1:
\begin{equation}\label{case1b}
\varepsilon_{1,2}<0, \quad \text{min}(D_{12},D_{21})<1
\end{equation}

\item The case of stability of non-trivial state off coordinate axes $S_4$ and
$M_4$ (Fig.\ref{modes}b,c, Fig.\ref{modes1}c,d).  For  system (\ref{eq:modelA2}) this situation occur in two
cases. The first situation appears if $\delta_{1,2}>0$ (when isolated
subpopulations tends to synchrony) and suppressive couplings are weak:
\begin{equation}
 \label{case2a-1}
\lambda_1=\frac{\delta_2}{2}\delta_2-\delta_1 d_{21}>0,\qquad
\lambda_2=\frac{\delta_1}{2}\delta_1-\delta_2 d_{12}>0.
\end{equation}
The states $S_{2,3}$ have eigenvalues $-\delta_1\delta_{1},\ \lambda_{1}$, and 
$-\delta_2\delta_2$, $\lambda_2$, respectively, and therefore are saddles.  
The origin
is an unstable node $(\delta_{1,2}>0)$ and therefore the state $S_4$ is an attractor (note
that $S_4$ always exists while (\ref{case2a-1}) holds). We call this situation
``case of weak suppressive couplings'' because (\ref{case2a-1}) can be written as
$d_{12}<\frac{\delta_1}{2}\frac{\delta_1}{\delta_2},\
d_{21}<\frac{\delta_2}{2}\frac{\delta_2}{\delta_1}.$ 

The second situation appears if one of the subpopulations 
approaches to the asynchronous state (negative $\delta$) while another group tends
to synchrony and has positive influence on the first subpopulation:
\begin{equation}\label{case2a-2}
\delta_1>0,\ \delta_2<0,\ \lambda_1>0\ \quad\text{or}\quad \delta_1<0,\ \delta_2>0,\
\lambda_2>0.
\end{equation}
Condition $\lambda_1>0$ is equivalent to
$d_{21}<\frac{\delta_2\delta_2}{2\delta_1}$, what means that coupling $d_{21}$
should be negative and absolute value of $d_{21}$ should be large enough to
maintain partially synchronous state inside the second cluster (positive influence).

For the system (\ref{eq:modelB2}) the situation of global stability
of $M_4$ can be produced by two types of conditions. The first case is that of
positive $\varepsilon_{1,2}$ and weak suppressive couplings:
\begin{equation}\label{case2b-1}
\varepsilon_{1,2} > 0,\ D_{12}<1,\ D_{21}<1.
\end{equation}
Another case of global stability of $M_4$ occurs if
\begin{equation}\label{case2b-2}
\varepsilon_1<0,\ \varepsilon_2>0,\ D_{12}>1,\ D_{21}<1\ \quad\text{or}\quad \varepsilon_1>0,\
\varepsilon_2<0,\ D_{12}<1,\ D_{21}>1.
\end{equation}
The latter case differs from the previous one only by the direction of the flow on lines
$\rho_{1,2}=0$ and the type of unstable points $M_1,M_2,M_3$ (Fig.\ref{modes1}d).

\item The case of competition between subpopulations
(Fig.\ref{modes}d,\ref{modes1}e).
 In model (\ref{eq:modelA2}) this type of behavior arises
when
\begin{equation}\label{case3a}
\delta_{1,2}>0,\ \lambda_1<0,\ \lambda_2<0.
\end{equation}
According to (\ref{case3a}) the points $S_{2,3}$ are stable, while $S_1$ is unstable node
and $S_4$ is a saddle. This case corresponds to the situation
of strong suppressive couplings
$$d_{12}>\frac{\delta_1}{2}\frac{\delta_1}{\delta_2},\quad
d_{21}>\frac{\delta_2}{2}\frac{\delta_2}{\delta_1}.$$ 

Competitive behavior in the system (\ref{eq:modelB2}) is
produced by positive $\varepsilon_{1,2}$ and strong suppressive couplings
between subpopulations:
\begin{equation}\label{case3b}
\varepsilon_{0}>0,\ D_{12}>1,\ D_{21}>1.
\end{equation}

\item The case of global stability of synchronous state of only one cluster
($S_2$,$M_2$) (Fig.\ref{modes}e,f,\ref{modes1}f,g,h).

In model (\ref{eq:modelA2}) only one group is synchronous in two cases.  The 
first trivial situation is similar to
conditions (\ref{case2a-2}) (when one group approaches
to asynchronous state while another one tends to synchrony) but in this case
the active group does not have sufficient positive coupling to maintain
synchronization in the asynchronous subpopulation (Fig.\ref{modes}f):
\begin{equation}\label{case4a-2}
\delta_1>0\ \delta_2<0\ \lambda_1<0\quad\text{or}\quad\delta_1<0\ \delta_2>0\
\lambda_2<0.
\end{equation}
Under conditions (\ref{case4a-2}) only one of the fixed points $S_2$ or $S_3$
exists
and $S_1$ is always unstable. 
The second case occurs at an asymmetric interaction of intrinsically
active clusters (isolated clusters tend to synchronous regime):
\begin{equation}\label{case4a-1}
\delta_{1,2}>0,\quad \text{and}\quad
\lambda_1<0\ \lambda_2>0\quad\text{or}\quad\lambda_1>0\ \lambda_2<0
\end{equation}
In other words, it appears when one coupling coefficient is strong enough to
 fully suppress the synchrony in the opponent, for example
$d_{21}>\frac{\delta_2}{2}\frac{\delta_2}{\delta_1}$, while another one is weak
or even non-suppressing $d_{12}<\frac{\delta_1}{2}\frac{\delta_1}{\delta_2}.$
In this case one can prove that $S_4$ does not exist, point $S_2$ is a stable
node, $S_3$ and $S_1$ are saddles. Thus all trajectories approach stable node
$S_2$ which corresponds to the synchronous state of the first group and
to the asynchronous state of the second one. 
Because of this on the plane $(\rho_1,\rho_2)$
always exists heteroclinic trajectory connecting saddle point $S_3$ and stable
equilibrium $S_2$ (red line in Fig.\ref{modes}e).

Global stability of point $M_2 (M_3)$  of system (\ref{eq:modelB2}) 
occurs in several different cases. The
first case is similar to the situation in the system (\ref{eq:modelA2}) at
conditions (\ref{case4a-2}) when one group tends to synchrony ($\delta_n>0$),
another one approaches trivial state ($\delta_m<0$) and synchronous group does
not have sufficient positive infuence to maintain synchronization in the
asynchronous group:
\begin{equation}\label{case4b-1}
\varepsilon_1<0,\ \varepsilon_2>0,\ D_{12}<1\quad\text{or}\quad\varepsilon_1>0,\
\varepsilon_2<0,\ D_{21}<1.
\end{equation}
Corresponding phase planes are presented in Fig.\ref{modes1}g,h.
Another case is that of positive $\varepsilon_{1,2}>0$ and asymmetric
couplings: 
\begin{equation}\label{case4b-2}
\varepsilon_{1,2}>0,\;
D_{12}>1\ D_{21}<1\ \quad\text{or}\quad D_{12}<1\ D_{21}>1.
\end{equation}
Under described above conditions (\ref{case4b-1}), (\ref{case4b-2}) it is easy
to show that only one stable fixed point $M_2(1,0)$ exists, so
all trajectories approach $M_2$. In the case of (\ref{case4b-2}) a
sequence of heteroclinic orbits connecting $M_2$ and $M_3$ (red lines in
Fig.\ref{modes1}f) appears.

\item The case of bistability of trivial and fully synchronous states
(Fig.\ref{modes1}i). 

In model (\ref{eq:modelB2}) this happens for negative $\varepsilon_{1,2}$ and
strong synchronizing couplings:
\begin{equation}\label{case5b}
\varepsilon_{1,2}<0,\ D_{12}>1,\ D_{21}>1.
\end{equation}

\item Periodic behavior (Fig.\ref{modes1}j).

In model (\ref{eq:modelB2}) periodic solutions can be observed.
Conditions 
\begin{eqnarray}\label{case6b}
D_{12}>1,\ D_{21}>1,\ \varepsilon_1\varepsilon_2<0
\end{eqnarray}
provide saddle type of points $M_{1-4}$ and existence of equilibrium $M_5$ with
imaginary 
eigenvalues
$\pm i\sqrt{\frac{\varepsilon_1\varepsilon_2(d_{12}-1)(d_{21}-1)}{4d_{12}d_{21}}}
.$ 
Because model (\ref{eq:modelB2}) can be rewritten as a Hamiltonian one, one has
a family of periodic orbits.
\end{enumerate}



\end{document}